\documentclass[%
superscriptaddress,
preprint,
 amsmath,amssymb, aps, 
]{revtex4-2}

\usepackage[plainpages=false,pdfpagelabels,colorlinks=true,linkcolor=red,urlcolor=blue,citecolor=blue,pdftitle={Title},pdfauthor={},pdfdisplaydoctitle=true,pdfduplex=DuplexFlipLongEdge]{hyperref}

\usepackage{physics}
\usepackage{braket}
\usepackage{bm}
\usepackage{subfigure}
\usepackage{graphicx}
\usepackage{xr}
\makeatletter
\newcommand*{\addFileDependency}[1]{%
  \typeout{(#1)}
  \@addtofilelist{#1}
  \IfFileExists{#1}{}{\typeout{No file #1.}}
}
\makeatother

\newcommand*{\myexternaldocument}[1]{%
    \externaldocument{#1}%
    \addFileDependency{#1.tex}%
    \addFileDependency{#1.aux}%
}

\myexternaldocument{supplementary_info}

\begin{document}

\title[]{Polariton Creation in Coupled Cavity Arrays with Spectrally Disordered Emitters}

\author{J T Patton}
\thanks{Equal contribution}
\affiliation{Department of Electrical and Computer Engineering, University of California, Davis, CA 95616, USA}
\author{V A Norman}
\thanks{Equal contribution}

\affiliation{Department of Electrical and Computer Engineering, University of California, Davis, CA 95616, USA}
\affiliation{Department of Physics and Astronomy, University of California, Davis, CA 95616, USA}

\author{E C Mann}
\affiliation{Department of Electrical and Computer Engineering, University of California, Davis, CA 95616, USA}
\affiliation{Department of Electrical Engineering, Harvard University, Cambridge, MA, 02138, USA}

\author{B Puri}
\affiliation{Department of Electrical and Computer Engineering, University of California, Davis, CA 95616, USA}

\author{R T Scalettar}
\affiliation{Department of Physics and Astronomy, University of California, Davis, CA 95616, USA}

\author{M Radulaski}
\email{mradulaski@ucdavis.edu}
\affiliation{Department of Electrical and Computer Engineering, University of California, Davis, CA 95616, USA}


\begin{abstract}
Integrated photonics has been a promising platform for analog quantum simulation of condensed matter phenomena in strongly correlated systems. 
To that end, we explore the implementation of all-photonic quantum simulators in coupled cavity arrays with integrated ensembles of spectrally disordered emitters. Our model is reflective of color center ensembles integrated into photonic crystal cavity arrays. Using the Quantum Master Equation and the Effective Hamiltonian approaches, we study energy band formation and wavefunction properties in the open quantum Tavis-Cummings-Hubbard framework. 
We find conditions for polariton creation and (de)localization under experimentally relevant values of disorder in emitter frequencies, cavity resonance frequencies, and emitter-cavity coupling rates.
To quantify these properties, we introduce two metrics, the polaritonic and 
nodal participation ratios, that characterize the light-matter hybridization and the node delocalization of the wavefunction, respectively. 
These new metrics combined with the Effective Hamiltonian approach prove to be a powerful toolbox for cavity quantum electrodynamical engineering of solid-state systems.
\end{abstract}


\maketitle

\section{Introduction}

Quantum simulation has attracted scientific attention since the early 1980s ignited by Richard Feynman's vision of the necessity of quantum mechanics in the modeling of natural phenomena \cite{feynman2018simulating}. 
Proposed implementations have included atomic, trapped ion, superconducting and photonic platforms \cite{altman2021quantum, georgescu2014quantum, cirac2012goals, trabesinger2012quantum, https://doi.org/10.1002/andp.201200261, noh2016quantum, PhysRevA.80.060301}.  
Here we focus on solid-state optical systems due to their potential for growth into large-scale commercial quantum simulators \cite{saxena2021photonic, aspuru2012photonic, luo2015quantum, hartmann2008quantum, saxena2023realizing}.

Nanophotonic cavities with integrated quantum emitters have served as a rich playground for exploring quantum optics phenomena in solid-state systems.
This includes demonstrations of weak \cite{vuvckovic2003enhanced} and strong \cite{reithmaier2004strong} cavity quantum electrodynamical (QED) coupling, photon blockade and photon-induced tunneling \cite{faraon2008coherent}, ultra-fast modulation of optical signals \cite{shambat2011ultrafast}, and more. 
The large dipole moment of quantum emitters, paired with (sub)wavelength scale optical mode volumes in photonic crystal cavities, give rise to high optical nonlinearities and light-matter state hybridization that creates polaritons. 
Polaritonic interactions in nanophotonic systems can be several orders of magnitude higher than those achieved in atomic systems. 
Such strong interaction  has been at the core of theoretical proposals for quantum state transfer \cite{almeida2016quantum, bose2007transfer}, as well as for the photonic simulation \cite{schmidt2009strong, hayward2012fractional} of Bose-Hubbard and fractional quantum Hall physics. 
Here, the system is made of an array of coupled cavities, each in the strong coupling regime of cavity QED, and described by the Jaynes-Cummings-Hubbard model. 
However, this model has been experimentally hard to achieve.

While progress toward the realization of coupled cavity arrays (CCAs) with embedded emitters has been made with quantum dots \cite{majumdar2012cavity, majumdar2012design}, the spectral disorder of these emitters has been a major roadblock to developing a large-scale resonant system. 
This problem is not present to such an extent with color center emitters, which are atomic defects in wide band gap materials. 
Recently, color center integration with nanocavities in diamond \cite{sipahigil2016integrated, zhang2018strongly} and silicon carbide \cite{lukin20204h, bracher2017selective} has been demonstrated in the weak cavity QED coupling regime. 
Though this regime is unsuitable for studies of polaritonic physics, proposals to demonstrate strong cavity QED regime have been presented with cavities integrating several ($M$) emitters, as opposed to a single emitter. 
Additionally, there has been renewed interest in disordered cavity QED systems with the discovery of phenomena like collectively induced transparency \cite{lei_many-body_2023}. 
Such systems are described by the Tavis-Cummings, rather than the Jaynes-Cummings model. 
Here, the collective coupling of emitters to the cavity effectively boosts the light-matter interaction rate by a factor of $\sqrt{M}$. 
Due to the small, but nonzero, spectral disorder of color centers, the collective strong coupling is possible within the \emph{cavity protection} regime, if its rate overcomes the spectral disorder $\Delta$ of color centers \cite{radulaski2017nonclassical, zhong2017interfacing}, i.e. $\Delta < g\sqrt{M}$. 
Such disordered multi-emitter cavity systems have been explored for applications in quantum light generation \cite{radulaski2017photon, trivedi2019photon, white2021superradiance}.

Here, we explore how all-photonic quantum simulators based on coupled cavity arrays can benefit from an increased interaction rate established in multi-emitter cavity QED. 
We expand the Jaynes-Cummings-Hubbard approach to the \textit{spectrally disordered} Tavis-Cummings-Hubbard model (TCHM) \cite{dull2021quality} and define conditions for polariton creation utilized in all-photonic quantum simulation, aided by the introduction of new localization metrics inspired by condensed matter approaches. 
Our model targets applications in technologically mature solid-state platforms and is reflective of the state-of-the-art parameters achieved in silicon carbide and diamond color center hosts. 
We find system limits that can be guiding for future experiments with polaritons in coupled cavity arrays:
1) polaritonic states are easier to create in systems where emitter/cavity interaction exceeds cavity hopping;
2) polariton creation in an array of lossy cavities can be achieved via integration of an increased number of emitters per cavity even in disordered  ensembles;
3) as in single cavities, disordered emitter ensembles in coupled cavity arrays can create polaritons by increasing the number of emitters per cavity to reach the cavity protection condition; and
4) disorder in resonances in a coupled cavity array localizes polaritons if the difference in frequencies between neighboring cavities exceeds the cavity hopping rate.

\begin{figure}[htb]
    \centering
    \includegraphics[width=0.8\linewidth,angle=0]{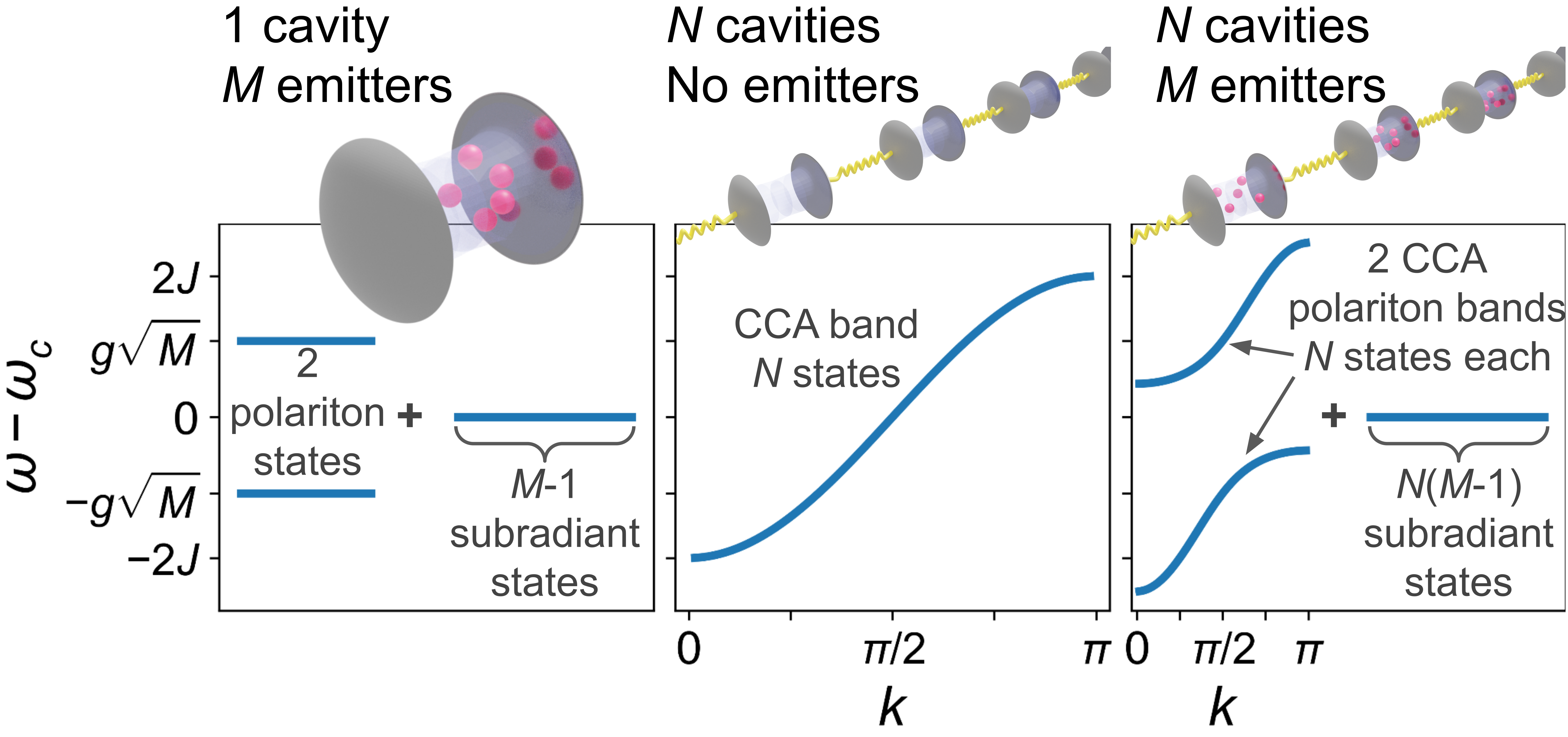}
    \caption{The eigenspectra of non-disordered coupled cavity arrays. (Left) One cavity with $M$ emitters has two polariton states and $M-1$ subradiant states. (Middle) CCA of $N$ cavities with no emitters has one CCA band of $N$ states. (Right) CCA of $N$ cavities and $M$ emitters per cavity has two polariton bands, upper and lower, of $N$ states each and $N(M-1)$ subradiant states.}
    \label{fig:nondisordered-eigenspectrum}
\end{figure}

\section{The CCA QED model}

Our CCA QED model captures the single-excitation regime of the spectrally disordered TCHM comprised of emitter-cavity localizing interactions and cavity-cavity delocalizing interactions:
\begin{equation}
\begin{aligned}
H_{\rm TCHM}= &\sum_{n=1}^N  \Big\{
 \omega^{\phantom{\dag}}_{c,n} a^\dag_n a^{\phantom{\dag}}_n +\sum_{m=1}^{M_n} \Big[ \omega^{\phantom{+}}_{e,n,m} \sigma_{n,m}^+ \sigma_{n,m}^-  
\\  
 &+ g^{\phantom{\dag}}_{n,m} (a_n^\dag \sigma_{n,m}^- + \sigma_{n,m}^+ a^{\phantom{+}}_n) \Big]
 -J^{\phantom{\dag}}_{n,n+1}(a_n^\dag a^{\phantom{\dag}}_{n+1}+a_{n+1}^\dag a^{\phantom{\dag}}_n)
\Big\},
\label{eq:HTCHM}
\end{aligned}
\end{equation}
where $N$ is the number of cavities in the array, $M_n$ is the number of emitters in the $n$-th cavity, $\omega_{c,n}$ and $a_n$ represent the angular frequency and the annihilation operator of the $n$-th cavity, $\omega_{e,n,m}$, $\sigma_{n,m}^-$ and $g_{n,m}$ correspond to the angular frequency, the lowering operator and the emitter-cavity coupling rate of the $m$-th emitter in the $n$-th cavity, $J_{n, n+1}$ is the photon hopping rate between the enumerated neighboring cavities. In this work we will assume $J_{n, n+1}=J$ as those parameters are set by the cavity design that is experimentally more controllable than the other parameters of the model \cite{majety2021quantum}.

\subsection{Non-disordered CCA QED model}

Before examining the spectral disorder effects in CCA QED, let us first address the energy spectrum in the fully resonant system of a linear array of coupled cavities with identical emitters. 
Here, the eigenenergy spectrum features two CCA polariton bands with $N$ states, each, and a degenerate set of $N(M-1)$ subradiant states as illustrated in Figure \ref{fig:nondisordered-eigenspectrum}. 
The polariton band states are parameterized by discrete momenta $k = k_p = \pi p /(N+1)$, ($p=1,2,3,.., N$) as
\begin{align}
E(k) =  \omega_c - J \, {\rm cos} \, k  
\pm \sqrt{ J^2 \, {\rm cos}^2 \, k + M g^2}
\,\,.
\label{eq:states}
\end{align}

Further discussion of the derivation of these equations for the fully resonant case is presented in Section 1 of the Supplementary Information. The origin of these spectral features can be decomposed to the QED and the CCA components. 
The resonant Tavis-Cummings model of $M$ emitters in $N=1$ cavity has the spectrum of two polaritons and $M-1$ degenerate subradiant states, while a CCA of $N>1, M = 0$ cavities has a single spectral band of $N$ photonic states. 
The spectrum of the resonant TCHM is a product of these components as seen in Figure \ref{fig:nondisordered-eigenspectrum}. This model has close analogs to the condensed matter models it aims to simulate. 

\subsection{Condensed matter analogs of TCHM}

A more thorough discussion of these analogs and their limitations is available in the Supplementary Information, but we can begin by looking at the TCHM system with identical emitters ($\omega_{e,n,m} = \omega_e$). 
Much of the derivation of eigenstates precisely parallels the calculation of the band structure of tight binding Hamiltonians commonly studied in condensed matter physics. 
For example, the case of no emitters ($M_n=0$) corresponds to the $d=1$ (Bose) Hubbard Model (HM)\cite{fisher1989}, and the case with a single emitter in each cavity ($M_n=1$) to the Periodic Anderson Model (PAM)\cite{rice1985}:
the hopping of photons between cavities is analogous to the conduction electrons which hybridize between different sites, and the photon-emitter coupling maps onto the hopping between the conduction electrons and localized orbitals which, like the emitters, do not have a direct intersite (intercavity) overlap. 
Thus, in the single excitation sector, and for one emitter per cavity, the models are {\it identical}.
In practice, the \textit{polaritonicity} (degree of light-matter hybridization) upon which we focus in the sections to follow, holds lessons for \textit{singlet formation} in which local and itinerant electrons become tightly intertwined in the PAM and Kondo Lattice Model.

\subsection{Experimentally-informed TCHM simulation parameters} 
\label{section:params}
The parameters of our model have been selected as representative of silicon carbide and diamond color center platforms.
Recent demonstrations of emitter-cavity interaction in photonic crystal cavities support rates of approximately $g/2\pi \sim$ 2-7.3 GHz \cite{evans2018photon, lukin20204h}, therefore, we chose a constant value of $g/2\pi =$ 5 GHz. While a variation in the coupling rate $g$ among emitters is likely to occur due to their variable positioning inside the electromagnetic mode, our prior work indicates that the collective emitter-cavity coupling still takes place \cite{radulaski2017nonclassical} at a well defined rate of $g_M=\sqrt{\sum_{m=1}^{M}g_m^2}$.
Therefore, keeping $g$ constant among the emitters should not take away from the overall phenomenology studied here.

The experimentally demonstrated cavity loss rates reach as low as $\kappa/2\pi \sim$ 15-50 GHz \cite{evans2018photon, lukin20204h}, while recent modeled designs  could reduce these values by at least an order of magnitude \cite{majety2021quantum}. 
With a slight optimism, we chose cavity loss rate of $\kappa/2\pi =$ 10 GHz.
Our recent designs of photonic crystal molecules indicate that coupled cavity hopping rates can be straightforwardly designed in the range 1 GHz $<J/2\pi<$ 200 GHz \cite{majety2021quantum}, thus spanning systems from the dominant cavity QED to the dominant photonic interaction character, represented in our choice of values $J/g = 0.1, 1, 10$. 

Fabrication imperfections may yield drifts in cavity resonant frequencies and hopping rates. 
The effect of this issue was studied in another platform where GaAs coupled cavity arrays were integrated with quantum dots \cite{majumdar2012design} and indicates that the coupling strength is an order of magnitude higher than the frequency and hopping rate perturbations. 
We apply this assumption in our model, maintaining that all cavities are mutually resonant and all rates $J$ are constant. 

The spectral inhomogeneity of emitters in fabricated devices, the main study of our model, has been characterized as $\Delta \sim$ 10 GHz for a variety of emitters in silicon carbide and diamond \cite{schroder2017scalable, babin2022fabrication}. 
We represent this parameter through its relation to the collective coupling rate $g\sqrt{M}$ in a cavity, spanning the spectral inhomogeneity across a range of values. 
The spectral disorder is implemented by sampling emitter angular frequency, $\omega_e$, from a Gaussian distribution $P( \omega_e)=\frac{1}{\sqrt{2\pi}\sigma}{\rm exp}\{-\frac{(\omega_e-\omega_c)^2}{2\sigma^2} \}$ centered at $\omega_c$ with a width of $\Delta=2\sigma$.
It is worth noting that the vibronic resonances are three orders of magnitude larger than the inhomogeneous broadening, for example 8.7 THz for the silicon vacancy in 4H-SiC \cite{shang2020local, bathen2021resolving}, therefore the phonon side band is not expected to play a part in the collective emitter-cavity coupling process. 
Emitter lifetime in color centers is usually in the 1-15 ns range \cite{majety2022quantum}, we select the value of the spontaneous emission from the color center, $\gamma/2\pi$ = 1/5.8 GHz, as representative. 
Due to $\gamma$ being the lowest rate in a color center-based cQED system, its minimal variations among emitters of the same species \cite{babin2022fabrication} affect the system only marginally, therefore we assume it has a constant value; this is representative of systems like atoms and color centers.

Lifetime- and nearly lifetime-limited emission of color centers has been demonstrated upon photonic integration \cite{babin2022fabrication, lukin2022optical, trusheim2020transform}. 
Due to this experimental advance, our model does not consider the dephasing terms, though such analysis may prove valuable with further development of integrated coupled cavity arrays. 

With these experimental constraints in mind, we believe our simulations will be directly relevant to future fabricated multi-emitter photonic crystal cavity chains.

\section{Effects of disorder on polariton formation}

It is in general computationally expensive to solve the Lindbladian master equation  to obtain exact simulation results \cite{manzano2020short}. As such, we are restricted to simulating only very small scale systems ($\sim 6$ elements total, an element being a single cavity or emitter) even in the low excitation regime. 
The results of these exact simulations are available in Sections 3 and 4 of the  Supplementary Information; the matrix form of the Hamiltonian that describes the system simulated is available in Section 2 of the Supplementary Information. 
On the other hand, the effective Hamiltonian ($H_{\rm EFF}$) uses the established non-hermitian effective approach to modeling Hamiltonians that are too resource intensive for the current state of the art classical computers to solve. 
Its approximation effectiveness is limited to the single-excitation regime, which is suitable for our exploration. 
Taking the effective Hamiltonian approach a step further, we also introduce the nodal and polaritonic participation ratios ($P_N$ and $P_P$, respectively). 
This method is derived from the condensed matter participation ratio metrics \cite{laflorencie2016quantum} used to quantify a system's localization properties. 
The $P_N$ and $P_P$ metrics are applied to eigenstates of the $H_{\rm EFF}$ to quantify the delocalization and light-matter hybridization of the wavefunction, respectively.

To access modeling of larger systems we develop a software package in Python \cite{TCH-Heff} that diagonalizes the Effective Hamiltonian in the approximate single-excitation regime
\begin{align}
H_{\rm EFF}= H_{\rm TCHM} &-\frac{i}{2} \sum_{n=1}^N  \Big\{ \kappa^{\phantom{\dag}}_{n} a^\dag_n a^{\phantom{\dag}}_n  +\sum_{m=1}^{M_n}\gamma^{\phantom{+}}_{n,m} \sigma_{n,m}^+ \sigma_{n,m}^-  \Big\},
\label{eq:HEFFH}
\end{align}
thus reducing the computational complexity from exponential to polynomial (cubic) in $N\times \left( M +1 \right)$ for the single-excitation regime. 
With this approximate method we numerically solve systems with hundreds of elements compared to the several using the exact QME approach. 
Note that, in contrast to QME, this method does not contain a pump term, meaning it is agnostic to the starting cavity and its diagonalization will provide all possible states, regardless of their wavelength overlap with the initial cavity. 

\subsection{The Participation Ratio Approach: Metrics for characterizing disorder}
\label{section:Metrics}
The node-by-node and element-by-element analysis required to examine each of the eigenstates found using $H_{\rm EFF}$ in the previous sections is lengthy and not suitable for the much larger systems we will be exploring. In order to efficiently analyze these much larger systems, we develop new metrics for the characterization of TCHM wavefunctions, inspired by practices in Condensed Matter Physics. 
The phenomenon of Anderson localization describes the loss of mobility of quantum particles  due to randomness \cite{anderson58absence}.
Originally studied in the context of non-interacting electrons hopping on a lattice with disordered site-energies, where all eigenstates were shown to be localized in  spatial dimension less than or equal to two \cite{abrahams1979scaling,lee85disordered}, Anderson localization has subsequently been extensively investigated in many further contexts, including the effect of interactions \cite{miranda05disorder}, correlations in the disorder \cite{croy2011anderson}, and importantly, new experimental realizations from cold atomic gases \cite{chen08phase,white09strongly} to transport in photonic lattices \cite{segev2013anderson}.

A useful metric for quantifying the localization of a wavefunction $v_p$, employed in these studies, is the participation ratio, $P = \left[\sum_{p}|v_p|^4\right]^{-1}$
 \cite{kramer1993localization} 
and its generalizations \cite{murphy2011generalized}. 

Instead of measuring the participation ratio among all $N(M
+1)$ vector components, we adapt $P$ to two new metrics that measure the participation among $N$ nodes (cavity-emitter sets), and two cavity- and emitter-like components. 
We define the nodal participation ratio,
\begin{align}
P_N =  \left[ \sum_ {n=1}^N \big( \, \braket{\mathcal{N}_{ph,n}} + \braket{\mathcal{N}_{e,n}} \, \big)^2 \, \right]^{-1},
\label{eq:PN}
\end{align}
where $\mathcal{N}_{ph,n} = a_n^\dag a^{\phantom{\dag}}_n$ and $\mathcal{N}_{e,n} = \sum_{m=1}^{M_n} \sigma_{n,m}^+\sigma_{n,m}^-$ are the usual number operators for each state $v_p$ representing cavity excitation and the sum of all emitter excitation in a cavity. 
Like the classic participation ratio, $P_N$ is at a minimum (maximum) when the wavefunction is localized (delocalized). 
Next, we define the polaritonic participation ratio, or \textit{polaritonicity},
\begin{align}
P_P =  \left[ \left( \sum_{n=1}^N   \braket{\mathcal{N}_{ph,n}}\right)^2 + \left(\sum_{n=1}^N \braket{\mathcal{N}_{e,n}} \right)^2 \right]^{-1},
\label{eq:PP}
\end{align}
which is minimized when the wavefunction has completely cavity-like or completely emitter-like character and is maximized for an equal superposition of cavity- and emitter-like components. 
Here we assume the character is polaritonic when there is any type of light-matter hybridization whether or not it is coming from the same node and therefore note that a wavefunction can be polaritonic even when the cavity and emitter excitations do not belong to the same node.

These two new metrics allow us to seamlessly characterize multi-emitter CCAs. 
We normalize the metrics to $1$ for easy comparison between the different model parameter cases.
To avoid numerical divide by zero errors, we set the identical emitter case of the leftmost column to have a small but nonzero value ($\Delta = \epsilon \approx 10^{-7}$).
\footnote{
 Since $\langle {\cal N}_{ph,1} \rangle =  
 \langle {\cal N}_{ph,2} \rangle = \frac{1}{4} $, and
 $\langle {\cal N}_{e,1} \rangle =  
 \langle {\cal N}_{e,2} \rangle =  
 \langle {\cal N}_{e,3} \rangle =  
 \langle {\cal N}_{e,4} \rangle = \frac{1}{8} $, 
 the explicit calculation from Eq.~\ref{eq:PN} is
 $P_N  = \big( (\frac{1}{4} + \frac{1}{8} +\frac{1}{8})^2 + 
 (\frac{1}{4} + \frac{1}{8} +\frac{1}{8})^2 \big)^{-1} = 2$.
 This is the maximal possible value and is then normalized
 to $P_N=1$.
 Likewise from Eq.~\ref{eq:PP},
 $P_P  = \big( (\frac{1}{4} +\frac{1}{4})^2 + 
 (\frac{1}{8} +\frac{1}{8} +\frac{1}{8} +\frac{1}{8})^2 \big)^{-1} = 2$.
 }

\subsection{Polaritonicity and localization as a function of the spectral disorder of the emitter ensemble}
\begin{figure}[htb]
    \centering
    \includegraphics[width=0.8\linewidth]{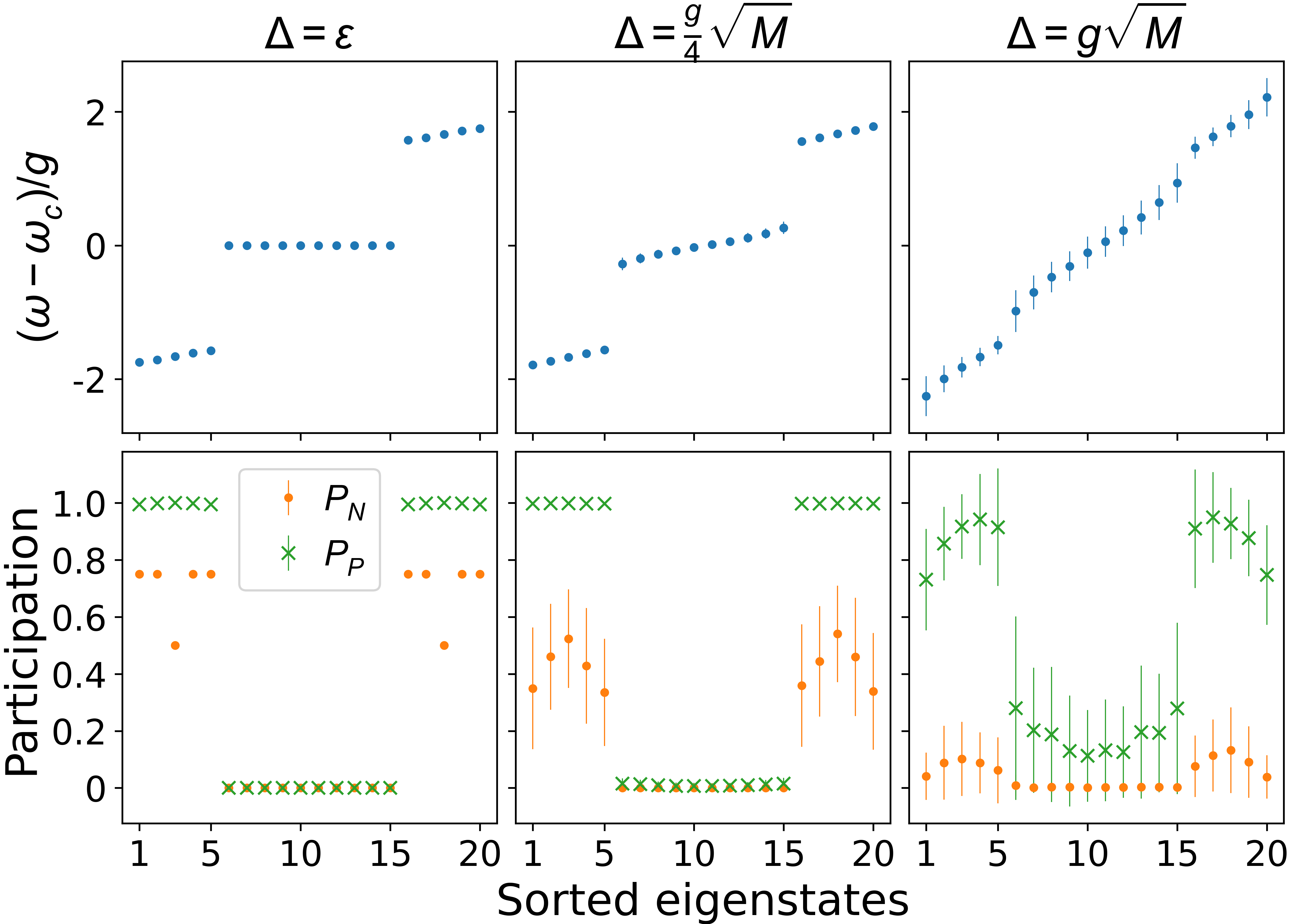}
    \caption{Effects of increasing disorder, $\Delta$, on the eigenstates of a system of $N=5$ cavities and $M=3$ emitters per cavity with $J/g=0.1$. (Top) Energy eigenvalues and (bottom) participation ratios $P_N$, $P_P$. Increasing $\Delta$ lifts the degeneracy of the subradiant states (flat middle band) decreasing the gap between the polaritonic bands and the subradiant states. Increasing $\Delta$ also causes an Anderson-like node localization in the polariton bands as shown by $P_N$ decreasing, but the polaritonic states remain mostly polaritonic. Mean of 100 random realizations; error bars are one standard deviation. $g$, $\kappa$, $\gamma$ are detailed in Section \ref{section:params}.
    }
    \label{fig:5x3smallJ}
\end{figure}

We investigate the effects of spectral disorder on large-scale TCHM systems with an open array of $N=5$ cavities with $M=3$ emitters per cavity on the localization and polaritonicity of the eigenstates of Eq.~\ref{eq:HEFFH}. In this section we set  $\omega_{c,n}=\omega_c$ and $g_{n,m}=g$. Figure \ref{fig:5x3smallJ} explores the regime where cavity QED dominates the photon hopping $J/g = 0.1$. 
For vanishing spectral disorder, $\epsilon$, the eigenspectrum has the shape resembling the features of Figure \ref{fig:nondisordered-eigenspectrum}: two highly polaritonic delocalized CCA bands with $N=5$ states and $N(M-1)=10$ highly localized subradiant states, suitably characterized by the polaritonic and nodal participation ratio values. 
For moderate disorder, the polaritonic properties of eigenstates are maintained, while the nodal localization somewhat increases for polaritonic band states. 
The degeneracy of the subradiant states is lifted and the spectral gaps diminish as we move into the strong disorder regime wherein $\Delta \approx g_M$, which is usually considered a cutoff for cavity protection.
Most states become highly localized, demonstrated by the significant drop in $P_N$ value and the subradiant states gain a cavity component, as quantified by the increase of the $P_P$ value.
While similar trends can be observed, the main difference is seen in the reduction of the number of polaritonic states in the CCA bands as the wavefunction obtains a higher cavity-like character.

\begin{figure}[htb]
    \centering
    \includegraphics[width=0.8\linewidth,angle=0]{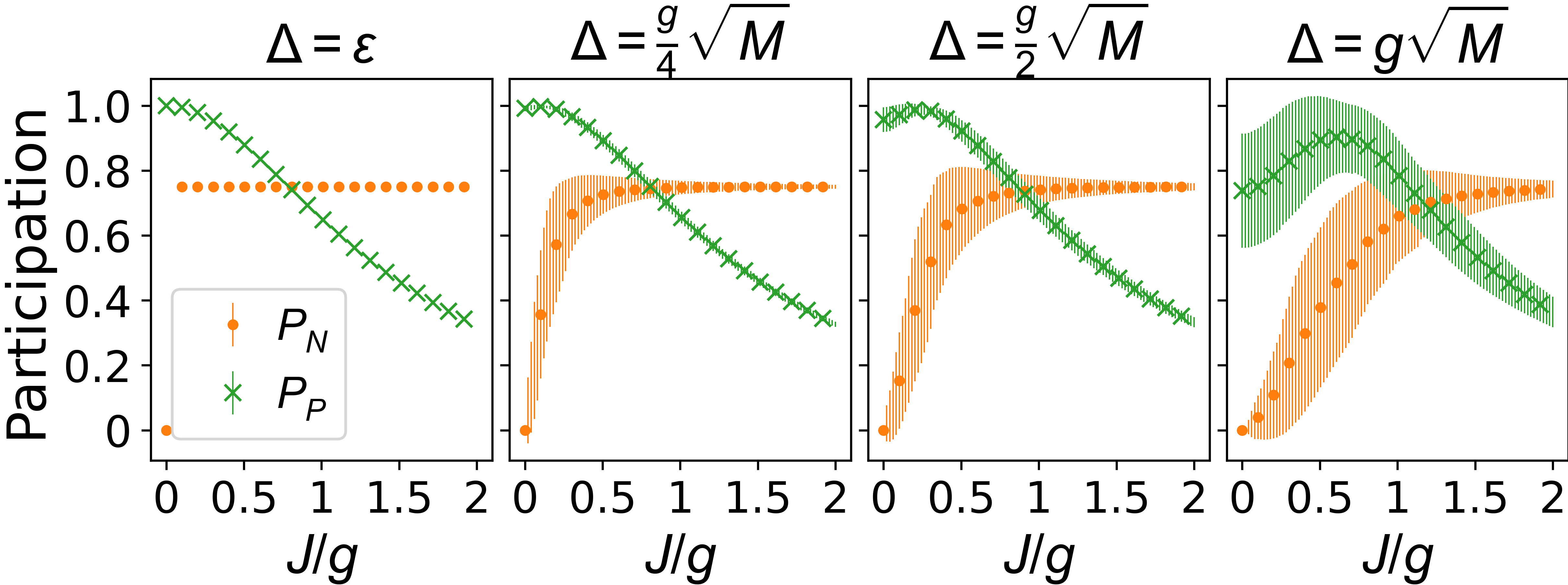}
    \caption{The nodal and the polaritonic participation ratios for the lowest energy eigenstate of a CCA with $N=5$ and $M=3$ for increasing $J/g$. A small amount of disorder causes the state to node localize for small $J/g$. Mean of 100 random realizations; error bars are one standard deviation. $N=5$, $M=3$; $g$, $\kappa$, $\gamma$ are detailed in Section \ref{section:params}.}
    \label{fig:5x3PcontJ}
\end{figure}

This brings us to look into the formation of a polaritonic state in spectrally disordered CCA QED as a function of an increasing $J/g$ ratio. Figure~\ref{fig:5x3PcontJ} shows the polaritonicity and localization of the lowest energy eigenstate.

When the photonic nature of the interaction increases, so does the cavity-like character of the wavefunction, reducing its level of polaritonicity. 
While this holds true for low and moderate values of disorder, in the case of high disorder, we observe an increase of $P_P$, before the decline. 
This is an artifact of the disorder which randomly modifies the nature of the lowest eigenstate in the system to be more emitter-like, until the interaction value increases to a level that offsets the issue. 
This trend is paired with the increase in the delocalization metric $P_N$ as the wavefunction loses the dominant emitter-like characteristic.

A decrease in polaritonicity and delocalization of the wavefunction take place for a range of system parameters. 
At low $J/g$ there is less variance in $P_P$ for a larger $\Delta$ compared to a higher photon hopping rate, suggesting that, as in the single node Tavis-Cummings model \cite{radulaski2017nonclassical}, the stronger cavity-emitter coupling compared to combined cavity losses (in the TCHM this includes cavity-cavity coupling) provides better cavity protection against the disorder in the TCHM.

\begin{figure}[htb]
    \centering
    \includegraphics[width=0.8\linewidth,angle=0]{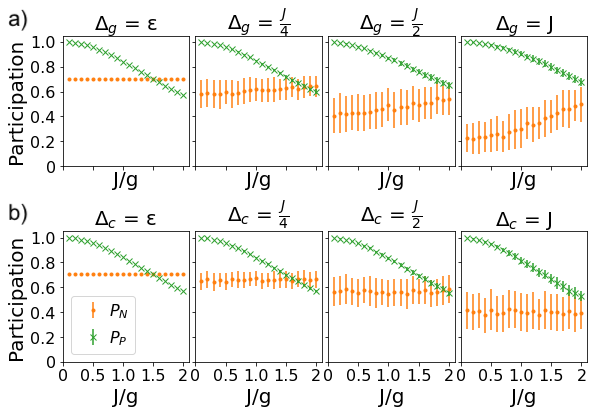}

    \caption{The nodal and polaritonic participation values for the lowest energy eigenstate of a CCA with $N=5$ and $M=10$ showing for increasing $J/g$ and increasing values of disorder in $g$ (top) and $\omega_c$ (bottom). 
    Means of  100 runs; error bars are one standard deviation. 
    In the top plots, $g_i/2\pi$ of each emitter is randomly taken from a Gaussian centered at $5$ MHz with standard deviation $\Delta_g /2$ 
    In the top plot, the coupling strength, $g/2\pi$, of each emitter is randomly taken from a Gaussian centered at $5$ MHz with FWHM $\Delta_g$ that is limited to values between zero and ten.
    In the bottom plots $\omega_c/2\pi$ of each cavity is pulled from a Gaussian centered at $0 \mathrm{MHz} = \omega_e$ with standard deviation $\Delta_c/2$. The nodal localization effects of increasing $\Delta_g$ and $\Delta_c$ are comparably less than those seen in Figure \ref{fig:5x3PcontJ} for increasing $\Delta_e$. }
    \label{fig:5_1}
\end{figure}

\begin{figure}[htb]
    \centering
    \includegraphics[width=\linewidth,angle=0]{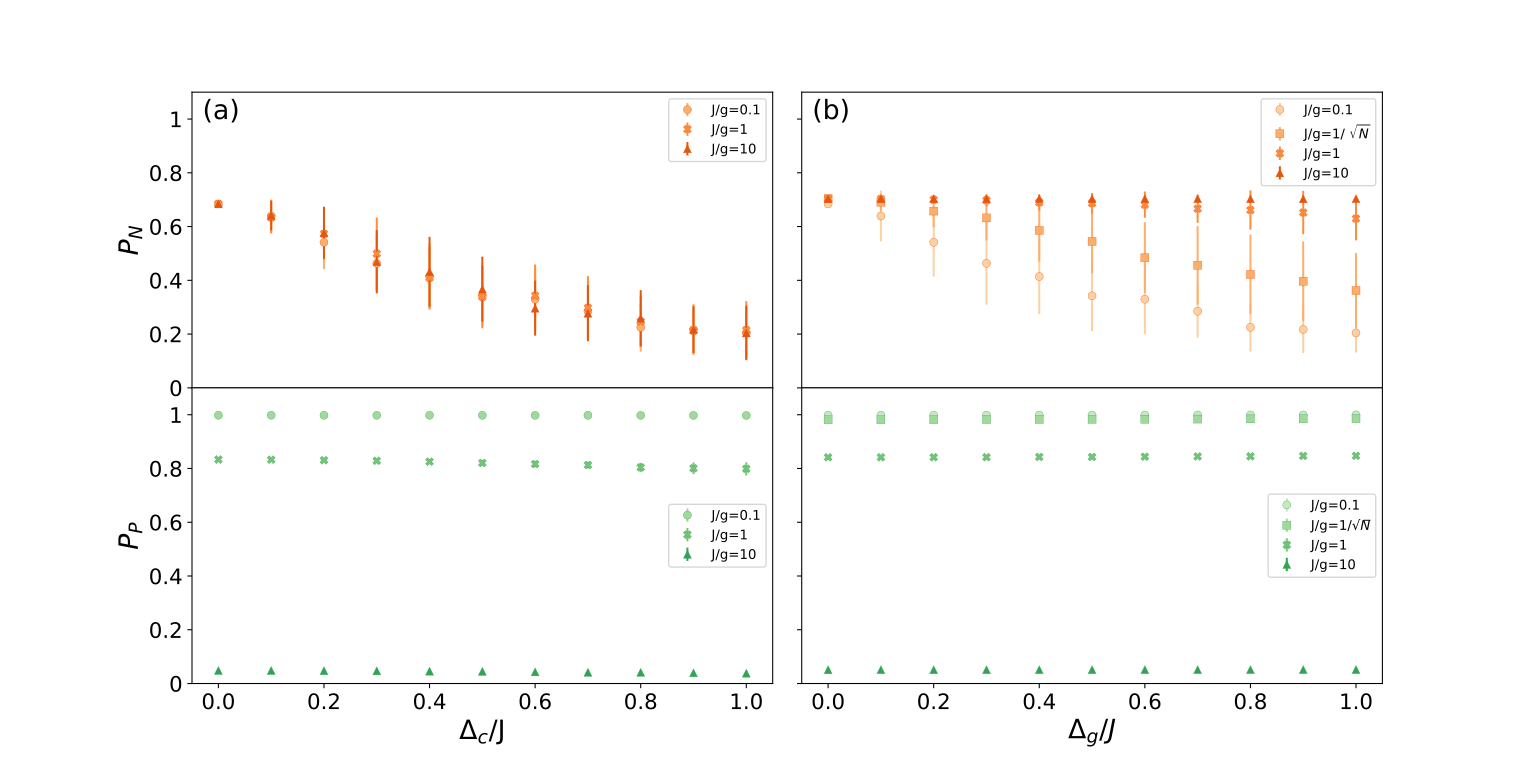}

    \caption{Plot (a) shows the average $P_N$ and $P_P$ for the lower energy eigenstate over 100 runs for systems with increasing $\Delta_c$. Plot (b) shows the same but for increasing $\Delta_g$. Parameters are the same as those reported in Figure \ref{fig:5_1}.  Various values of $J/g$ are denoted by the different plot markers in each plot.}
    \label{fig:6_1}
\end{figure}
\subsection{Other experimentally relevant forms of disorder} 
While we expect disorder due to spectral inhomogeneity of the emitters to dominate the creation of polaritons in the TCHM, we explore the effects of other potential sources of disorder that will arise in experimental implementations of these systems.

One potential source is the disorder of the emitter-cavity coupling rate, $\Delta_g$.
This disorder is important experimentally to consider since the dipole direction and the exact placement and number of color centers relative to the cavity electromagnetic field distribution cannot be precisely controlled in implantation. 
Effectively, collective coupling, $G_n = \sqrt{ \sum_{m=1}^{M_n} |g_{n,m}|^2 }$ \cite{radulaski2017nonclassical}, at each node can vary according to the statistically varying number of emitters or their positioning within the mode. Figure \ref{fig:5_1}a (top) suggests that increasing $\Delta_g$ leads to states becoming localized (i.e.~$P_N$ decreases) as seen in the global decrease of $P_N$ as $\Delta_g$ increases.
This is born out in Figure \ref{fig:6_1}b (bottom) in which, for moderate values of $J$, $P_N$ decreases as $\Delta_g$ increases. 
Not surprisingly, this localization effect is counteracted by large values of $J$ as shown by the modest upward trend in $P_N$ as $J/g$ approaches $2$. 
Potentially surprising, however, is the trend seen most clearly in Figure \ref{fig:6_1}b (bottom); the polaritonicity of the lowest energy eigenstate does not change as $\Delta_g$ is increased, but instead remains constant for a fixed value of $J$. 

A second experimentally significant source of disorder is the variation in the cavity frequency from one cavity to the next, $\Delta_c$. 
Any minor variance in nanofabrication from one cavity to the next will alter $\omega_c$ \cite{saxena2023realizing}. 
The decrease in $P_N$ as disorder in $\omega_c$ increases in Figure \ref{fig:5_1}b (top) suggests that for an increasing cavity frequency disorder, the lowest energy eigenstates become more localized. 
In fact, Figure \ref{fig:6_1}a (top) suggests that there will always be a drop in $P_N$ as $\Delta_c$ increases and that regardless of the strength of the cavity-cavity coupling, the rate at which $P_N$ decreases is the same. 
On the other hand, Figure \ref{fig:6_1}b (bottom), like Figure \ref{fig:6_1}a (bottom), suggests that the polaritonicity of the system is fairly tolerant of cavity fabrication errors.

Both $\Delta_c$ and $\Delta_g$ have similar effects on $P_P$, the scale of which is set by the value of $J$. In contrast to variations in $\omega_e$ that effect both $P_P$ and $P_N$, variations in $\omega_c$ and $g$ affect only the localization properties of polaritons. In practice, this means that if neighboring cavities in the CCA have sudden jumps in $\omega_c$ or $g$ it can effectively cut the CCA into two smaller CCAs with independent TCHM physics from one another.

\section{Discussion}

In this work we characterize the CCA QED eigenstates described by the Tavis-Cummings-Hubbard model. 
Our goal is to provide a guiding tool for experimental implementations through the engineering of the CCA parameters.
 
Using the new participation ratio metrics, inspired by condensed matter physics studies of localization and band mixing, we confirm that highly polaritonic states can be formed in coupled cavity arrays despite the presence of spectral disorder in emitter ensembles and quantify the cavity protection effect. 
While the systems with a dominant cavity QED interaction, relative to the photon hopping rate, support creation of numerous polaritonic states, we find that other parts of the parameter space can also be utilized to study polaritonic physics.

We suggested approximate analogies between the case of $M_n=1$ emitter in each cavity with the periodic Anderson model where a single $f$ orbital on each site hybridizes with a conduction band, and the Kondo lattice model where the local degree of freedom is spin-1/2. 
Condensed matter systems which connect to the multi-emitter case $M_n>1$ also have a long history, both in the investigation of multi-band materials and also as a theoretical tool providing an analytically tractable \textit{large-$N$} limit \cite{bickers87,bickers87rev}. 
Indeed, large-$N$ systems, realized for example by alkaline earth atoms in optical lattices, are also at the forefront of recent work  in the atomic, molecular and optical physics community \cite{taie12,ozawa18,ibarra21}.
In short, the TCHM offers a context to explore intertwined local and itinerant quantum degrees of freedom which, while distinct from condensed matter models, might still offer insight into their behavior.

These TCHM systems can ostensibly be realized in a number of photonic frameworks, from atoms in mirrored cavities, to quantum dots in nanophotonics. 
It is difficult, however, to experimentally create atom-based systems that couple multiple cavities together and to create large numbers of quantum dots that emit within the relatively modest range of disorder that we have shown will recreate polariton dynamics. 
As such, the most likely experimental realization of our systems will be in color center based nanophotonics.

\vskip0.20in \noindent
\textbf{Acknowledgements}
M.R. and V.A.N. acknowledge support by the National Science Foundation CAREER award 2047564. R.T.S.~was supported by the grant DE‐SC0014671 funded by the U.S. Department of Energy, Office of Science. M.R., V.A.N. and R.T.S. M.R. and R.T.S. acknowledge support by the University of California Multicampus Research Programs and Initiatives (CIRQIT pilot award).

\newpage

\end{document}


\Large{\bf{Supplementary Information}}

\label{section:SuppInf}
{}
\normalsize
\section{Condensed Matter analogies}
The resonant cavity-emitter arrays 
($\omega_{c,n}=\omega_c$, 
$\omega_{e,n,m}=\omega_e$, 
$g_{n,m}=g$, 
$J_{n,n+1}=J$),
with periodic boundary conditions have a closed form solution for the eigenvalues and eigenvectors.
Much of the derivation precisely parallels the calculation
of the band structure of tight binding Hamiltonians. The case of 
no emitters ($M_n=0$) corresponds to the $d=1$ (Bose) Hubbard Model (HM), and the
case with a single emitter in each cavity ($M_n=1$) to the Periodic Anderson Model (PAM).
Here we review those results with an emphasis on the condensed matter analogies.

In the case $M_n=0$ one can diagonalize $H_{\rm TCHM}$ by introducing momentum 
creation and  destruction operators,
\begin{align}
a_k^{\dagger} = \frac{1}{\sqrt{N}} \, \sum_{l=1}^{N} e^{i k l} a_l^{\dagger} 
\hskip0.30in
a_k^{\phantom{\dagger}} = \frac{1}{\sqrt{N}} \, \sum_{l=1}^{N} e^{-i k l} a_l^{\phantom{\dagger}}
\end{align}

The transformation is canonical. $a_k^{\dagger} $ and
$a_k^{\phantom{\dagger}}$ obey the same bosonic commutation relations as the original real space operators.
$H_{\rm TCHM}$ is diagonal,
\begin{align}
H_{\rm TCHM} = \sum_k \big( \omega_c - 2 J \, {\rm cos} \, k \big) \,
a_k^{\dagger}  a_k^{\phantom{\dagger}}
\label{eq:TCHHUB}
\end{align}
The eigenenergies are,
\begin{align}
E(k) =  \omega_c - 2 J \, {\rm cos} \, k  \,\,.
\label{eq:TCHHUBEk}
\end{align}
The momenta are discrete, $k = k_p = 2 \pi p /N$ with $p=1,2,3, \cdots N$.
In the thermodynamic limit the $E(k)$ form a continuous band with a density of states
that diverges at the band edges $E(k) = \omega_c \pm 2J$.

In this no-emitter limit, Eq.~\ref{eq:TCHHUB} actually provides the solution for any number of
excitations.  The many-excitation energies are just sums of the single particle $E(k)$ subject to
the photon indistinguishability implied by the commutation relations.  This solubility of the 
many excitation system is unique to the no-emitter limit, as discussed further below.

In the case $M_n=M=1$ one can again solve for the eigenvalues of $H_{\rm TCHM}$ 
by going to momentum space, but only in the single excitation sector.  The reason
is that the photon and emitter operators do not obey a consistent set of
commutation relations.  Thus, even though it might appear that 
$H_{\rm TCHM}$ 
is soluble since it is quadratic in the operators,
it is therefore not possible to do the same sort of
canonical transformation to diagonalize.
It is most straightforward to define a set of single excitation states which
form a basis for the space, and then examine the matrix which arises from
the application of
$H_{\rm TCHM}$ 
to each one.  For simplicity, we focus on the \textit{resonant} case 
where $\omega_c=\omega_e \equiv \omega_0$, but the formulae are straightforward to generalize.
The energy eigenvalues are
\begin{align}
E(k) =  \omega_0 - J \, {\rm cos} \, k  
\pm \sqrt{ J^2 \, {\rm cos}^2 \, k + g^2}
\,\,.
\label{eq:TCHPAMEk}
\end{align}
The (unnormalized) eigenvectors are,
\begin{align}
\Psi_{\pm}(k) =  
\left(
\begin{array}{c}
 g   \\[0.2cm]
 J {\rm cos}\,k \mp \sqrt{J^2\,{\rm cos}^2\,k+g^2} \\
\end{array}
\right)
\label{eq:TCHPAMEv}
\end{align}
so that, while all states are polaritons in the sense of mixing photon and emitter
components, the relative weights depend on momentum $k$ and band index $\pm$.

This $M_n=M=1$ case easily generalizes to larger $M$.  There are again two polariton
bands, but with an enhanced photon-emitter hybridization $\sqrt{M} g$,
\begin{align}
E(k) =  \omega_0 - J \, {\rm cos} \, k  
\pm \sqrt{ J^2 \, {\rm cos}^2 \, k + M g^2}
\,\,.
\label{eq:TCHPAMEkM}
\end{align}
with a similar $g^2 \rightarrow M g^2$ change to the eigenvectors
of Eq.~\ref{eq:TCHPAMEv}.
The remaining $M-1$ bands have purely emitter components, and
are dispersionless, $E(k)=\omega_0$.

\begin{figure}[htb]
\centering
\includegraphics[width=0.8\linewidth]{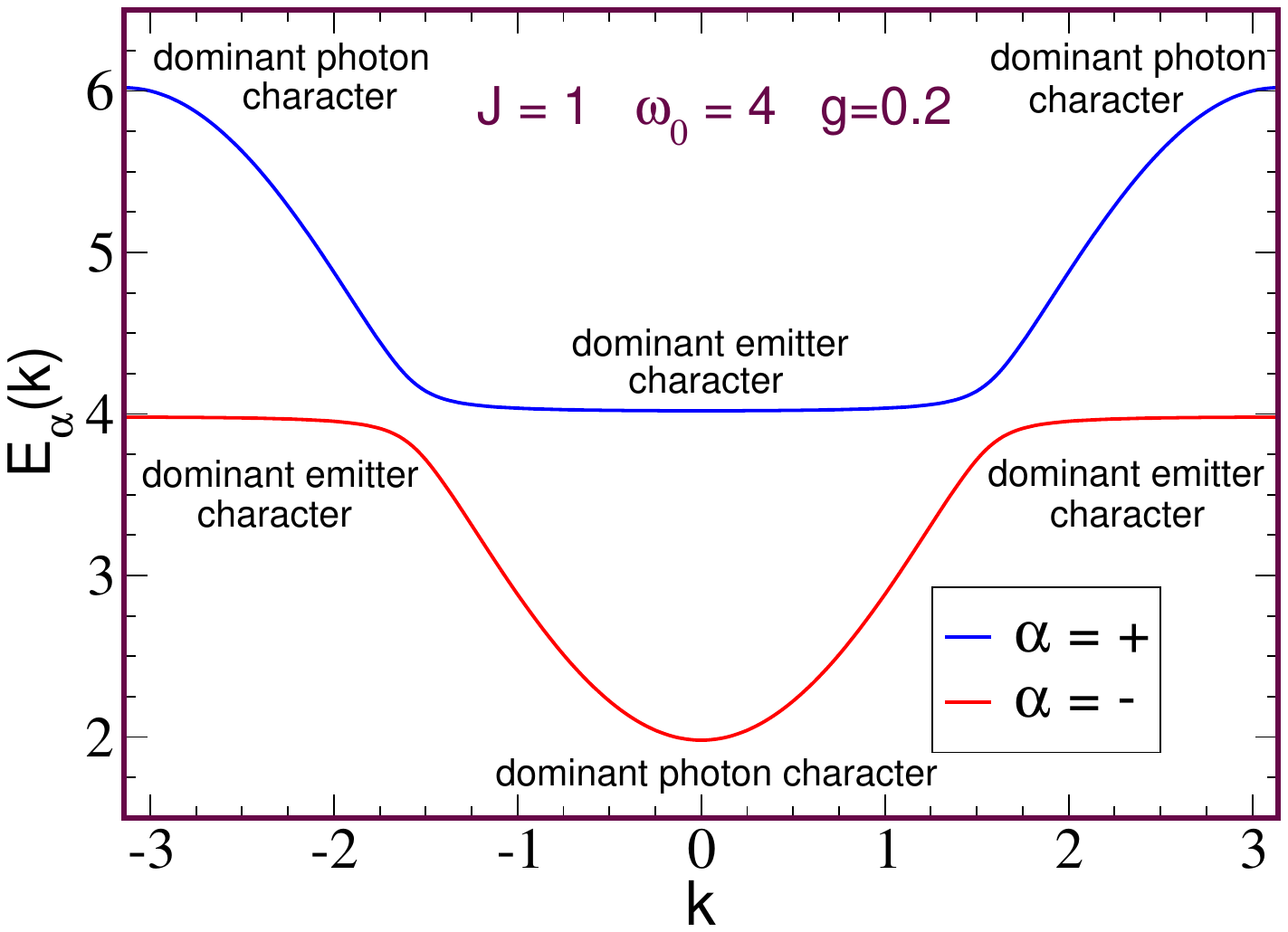}
\caption{The two polariton bands of a cavity-emitter system for 
representative parameters $\omega_c=\omega_e\equiv\omega_0=4$, cavity hopping $J=1$, and cavity-emitter coupling $g=0.2$.
If the number of emitters $M>1$ the gap between the two bands is enhanced from $\Delta=2g$ to $\Delta=2\sqrt{M}g$ and there are $M-1$ additional \textit{flat} emitter bands at $E_\alpha(k) = \omega_0$ in the gap between the two polariton bands.}
\label{fig:PAMbands}
\end{figure}

A useful approximate visualization of the eigenvalues and eigenvector weights 
is provided by 
drawing the bands as in Figure~\ref{fig:PAMbands}.  For $g=0$ the photons
have $E(k) = \omega_0 - 2J \, {\rm cos}k$ and the emitters $E(k)=\omega_0$.
When these two energy levels are hybridized by $g$ there is a level repulsion
at their crossing point at $k=\pm \frac{\pi}{2}$ and an energy gap $\Delta = 2g$
opens.  The relative photon-emitter compositions of the states can be 
inferred from the degree to which the polariton energy matches one of the
initial ($g=0$)  photon or emitter bands.  Polariton energies which
are close to the original flat $E(k) = \omega_0$ emitter band are dominantly
made up of emitter excitations, while those close to the original dispersing
$\omega_0 - 2J \, {\rm cos}k$ photon band are dominantly cavity excitations.

\section{Matrix form of $H_{TCHM}$}

\setcounter{MaxMatrixCols}{20}
As an illustrative example, the Hamiltonian of the largest system simulable on an M2 Macbook Pro with 32 GB of memory using the quantum master equation approach (2 cavities with 2 emitters in each cavity) is presented in matrix form:

\begin{equation}
    \begin{bmatrix}
    \omega_{c,1} & g_{1,1} & g_{1,2} & J_{1,2} & 0 & 0\\
    g_{1,1} & \omega_{e,1,1} & 0 & 0 & 0 & 0 \\
    g_{1,2} & 0 & \omega_{e,1,2} & 0 & 0 & 0\\
    J_{1,2} & 0 & 0 & \omega_{c,2} & g_{2,1} & g_{2,2}\\
    0 & 0 & 0 & g_{2,1} & \omega_{e,2,1}& 0\\
    0 & 0 & 0 & g_{2,2} & 0 & \omega_{e,2,2}\\

    \end{bmatrix}
\end{equation}
For convenience the equation form is reproduced here:
\begin{equation}
\begin{aligned}
H_{\rm TCHM}= &\sum_{n=1}^N  \Big\{
 \omega^{\phantom{\dag}}_{c,n} a^\dag_n a^{\phantom{\dag}}_n +\sum_{m=1}^{M_n} \Big[ \omega^{\phantom{+}}_{e,n,m} \sigma_{n,m}^+ \sigma_{n,m}^-  
\\  
 &+ g^{\phantom{\dag}}_{n,m} (a_n^\dag \sigma_{n,m}^- + \sigma_{n,m}^+ a^{\phantom{+}}_n) \Big]
 -J^{\phantom{\dag}}_{n,n+1}(a_n^\dag a^{\phantom{\dag}}_{n+1}+a_{n+1}^\dag a^{\phantom{\dag}}_n)
\Big\},
\label{eq:HTCHM}
\end{aligned}
\end{equation}
The $\omega_{c,n} a^\dag_{n} a_{n}$ term gives the energy of the $nth$ cavity, similarly the $\omega_{e,n,m} \sigma_{n,m}^\dag \sigma_{n,m}$ term gives the energy of the $mth$ emitter in the $nth$ cavity. The interaction between this emitter and cavity is described by the coupling term: $g_{n,m} \left( a_{n}^\dag \sigma_{n,m} + \sigma_{n,m}^\dag a \right)$. Finally, the interaction between nearest neighbor cavities is given by the last term, $J_{n+1,n} \left( a_{n+1}^\dag a_{n} + a_{n}^\dag a_{n+1} \right)$.

\section{Exact simulation results}

\begin{figure}[htb]
        \includegraphics[height=3.0in,width=3.0in,angle=0]{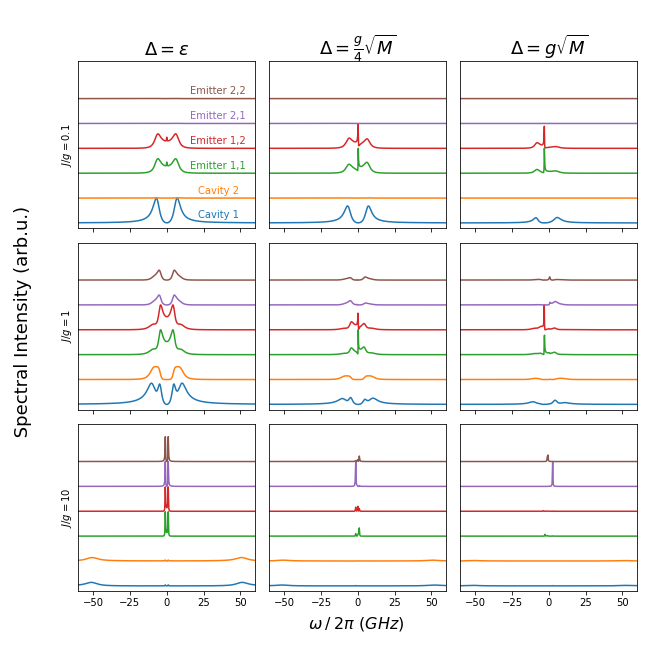} \label{fig:cav1pumpall}
        \includegraphics[height=3.0in,width=3.0in,angle=0]{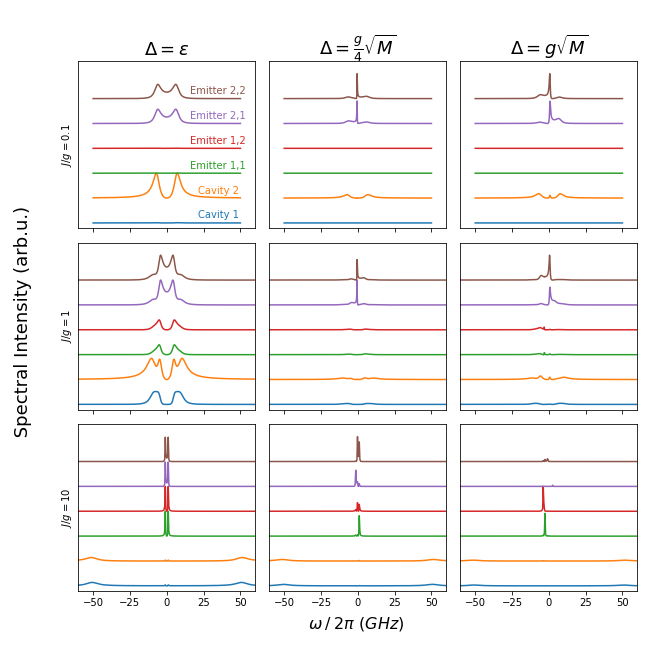} \label{fig:cav2pumpall}
    \caption{The top plots show the pump operator acting on Cavity 1 and the bottom set show the pump term acting on Cavity 2. Each line within a subplot refers to the transmission spectrum acquired by probing a particular element of the systems, either Cavities or any of the Emitters. Cavity 1 (blue) is coupled to Emitters 1,1 and 1,2 (green and red respectively). Cavity 2 is coupled to Emitters 2,1 and 2,2 (purple and brown respectively). The rows correspond to different cavity-cavity to cavity-emitter coupling ratios, with $J/g = 0.1$ (top row) having the least coupling between cavities, and $J/g = 10$ (bottom row) having the most. Moving from left to right increases the amount of disorder introduced via emitter emission wavelength detuning from a nominally zero amount ($\epsilon$), to detuning approximately equal to the cavity-emitter coupling ($g$).}
    \label{fig:cavpumpall}
\end{figure}

\begin{figure}[htb]
    \centering
    \includegraphics[width=\linewidth,angle=0]{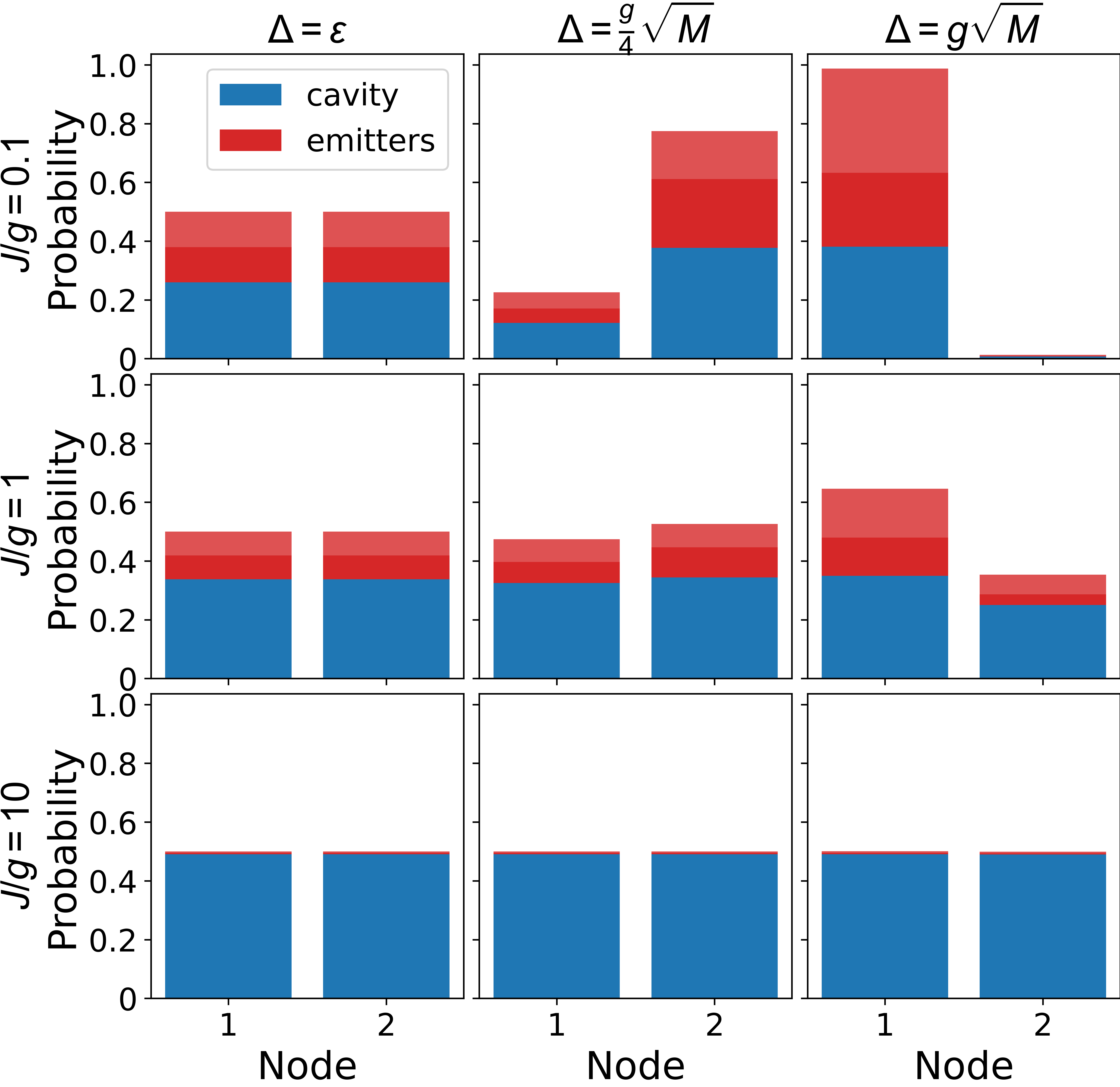}
    \caption{Node occupancy of the lowest energy eigenstate of a system with $N=2$ cavities and $M=2$ emitters per cavity with random $\omega_e$ sampled from a Gaussian distribution of width $\Delta$. Single random realization for each $\Delta$. $g$, $\kappa$, $\gamma$, and emitter frequencies are the same as those in Figure \ref{fig:cavpumpall}.}
    \label{fig:2x2v}
\end{figure}

We have developed a software package \cite{TCH-QME} that solves the Quantum Master Equation (QME). Our code uses the Quantum Toolbox in Python (QuTiP) \cite{qutip2} which solves the Lindbladian
\begin{align}
    \mathcal{L} \rho( t ) &= -i\big[H_{\rm TCH} ,  \rho(t)\big] 
    \nonumber \\
    &+  \sum_{n=1}^{N} \big\{ \frac{\kappa_n}{2} \mathcal{D}\left[ a_n \right]\rho(t) + \sum^M_{m = 1} \frac{\gamma_{n,m}}{2} \mathcal{D}\left[ \sigma_{n,m}^- \right] \rho(t) \big\} 
        \nonumber \\
    &+ P \mathcal{D}\left[ a^\dagger_1\right]\rho(t)
    \label{eq:QMTCH}
\end{align}
where $\mathcal{D}\left[c\right] \rho(t) = 2 c \rho(t)c^\dag - c^\dag c \rho(t) - \rho(t)c^\dag c$, $\kappa_n=\kappa$ is the cavity linewidth of the $n$-th cavity, $\gamma_{n,m}=\gamma$ is the emission rate of the $m$-th emitter in the $n$-th cavity, and $P$ is the optical (laser) pumping term. The spectral intensity reported in Figure \ref{fig:cavpumpall} is calculated as the Fourier Transform of the correlation function $\langle A^\dag (t + \tau ) A(t)\rangle$:
\begin{equation}
    S(\omega) = \int^\infty_{-\infty} \lim_{t \to \infty} \langle A^\dag(t + \tau) A(t)\rangle e^{-i\omega \tau} \,d\tau,
\end{equation}
where $A$ is replaced by the cavity annihilation and the emitter lowering operators, $a_n$ and $\sigma_{n,m}^-$.

The QME requires the use of the full density operator because of the non-number conserving term $2 c \rho(t)c^\dag$.

The density operator requirement makes solving this system highly resource-intensive (exponential in $N\times M$), as such we have restricted our exact calculations to small systems of six elements, or two coupled cavities with two emitters per cavity. Systems with more processing power may be able to simulate larger systems. Quantum trajectory theory is another potential method that may simulate larger systems using the quantum master equation approach.

We numerically solve the quantum master equation for a coupled 2 cavity system with each cavity also coupling to two emitters each. The results are shown in Figure \ref{fig:cavpumpall}. Along the top row, which shows the smallest inter-cavity coupling simulations, we see the most nodal localization regardless of emitter dispersion, as shown by only having peaks in Cavity 1 and Emitters 1 and 2 when pumping Cavity 1 and only having peaks in Cavity 2 and Emitters 3 and 4 when pumping Cavity 2. We can also see that the small inter-cavity coupling leads to highly polaritonic systems by comparing the peak heights of each cavity to those of its emitters and finding the ratios are close to unity (0.78 - 0.9). The largest inter-cavity coupling, $J/g = 10$ is heavily cavity-like as shown by cavity peak to emitter peak ratios of around 50. The middle row, $J/g = 1$ is partially polaritonic and partially photonic with peak ratios in the range 0.3-0.6. 
By introducing nonzero emitter detuning, we are able to access subradiant states for each of the three detuning values. These states are identified by the zero-width peaks found between the polariton peaks. 
Subradiant states are states that decay much slower than the polariton peaks and a number of proposals exist for their use in quantum information technologies including, light storage \cite{Facchinetti2016StoringLW} and quantum light generation \cite{radulaski2017photon}.

\section{Benchmarking of the Effective Hamiltonian approximation and the participation ratio metrics}

To benchmark the approximate Effective Hamiltonian approach against the exact Quantum Master Equation solution, we simulate the systems with same parameters in both models. Figure \ref{fig:2x2v} shows the lowest energy wavefunction calculated with the Effective Hamiltonian approach, corresponding to the excitations of the lowest energy peaks in Figure \ref{fig:cavpumpall}. The case of the vanishing spectral disorder shows equal contribution of Cavities 1 and 2, which corresponds to the identical looking plots of the Cavity 1 and Cavity 2 excitations in the QME spectra when Cavity 1 and Cavity 2 are pumped, respectively. The asymmetry of the nodal occupations in Figure \ref{fig:2x2v} for non-vanishing disorder (especially for $J/g=0.1$) matches the non-identicality of the Cavity 1/2, as well as Emitter 1.1,1.2 and Emitter 2.1,2.2 spectra in Figure \ref{fig:cavpumpall}. Localization of the wavefunction for an increasing disorder and $J/g \leq 1$ follow the exact solution trends described in the previous section.

\begin{figure}[htb]
    \centering
    \includegraphics[width=\linewidth,angle=0]{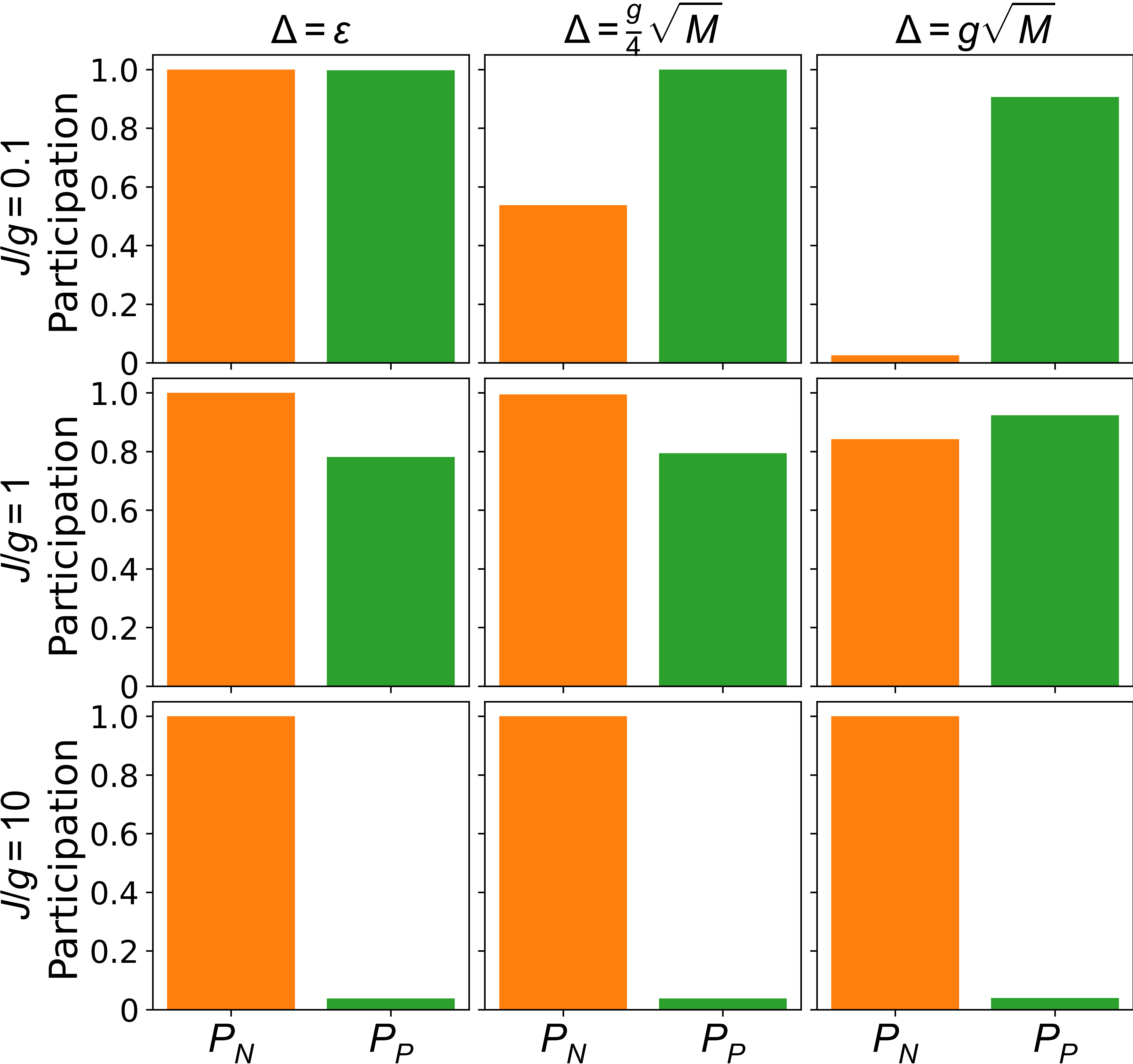}
    \caption{Participation ratios of the lowest energy eigenstate for a system with $N=2$ cavities and $M=2$ emitters per cavity with random $\omega_e$ sampled from a Gaussian distribution of width $\Delta$. Single random realization for each $\Delta$. $g=5$, $\kappa=10$, $\gamma=1/5.8$. The emitter frequencies are the same as those in Figure \ref{fig:cavpumpall}.}
    \label{fig:2x2P}
\end{figure}

These parallels are closely described by the nodal $P_N$ and the polaritonic $P_P$ participation ratio shown in Figure \ref{fig:2x2P}. The value of $P_N$ (orange) is maximized for a fully node-delocalized wavefunction, and reduces as the wavefunction tends to increasingly excite one cavity and its emitters with an increasing disorder. The value of $P_P$ (green) is maximized for the wavefunctions that have equal excitation distribution between cavities and the emitters, and reduces with an increasing photonic interaction (high $J/g$ value) as the cavities become predominantly excited.

We conclude that the Effective Hamiltonian and the participation ratio metrics are suitable for studies of the Tavis-Cummings-Hubbard model.

\section{Effects of spectral disorder on the TCHM energy spectrum}

\begin{figure}[htb]
    \centering
    \includegraphics[width=\linewidth,angle=0]{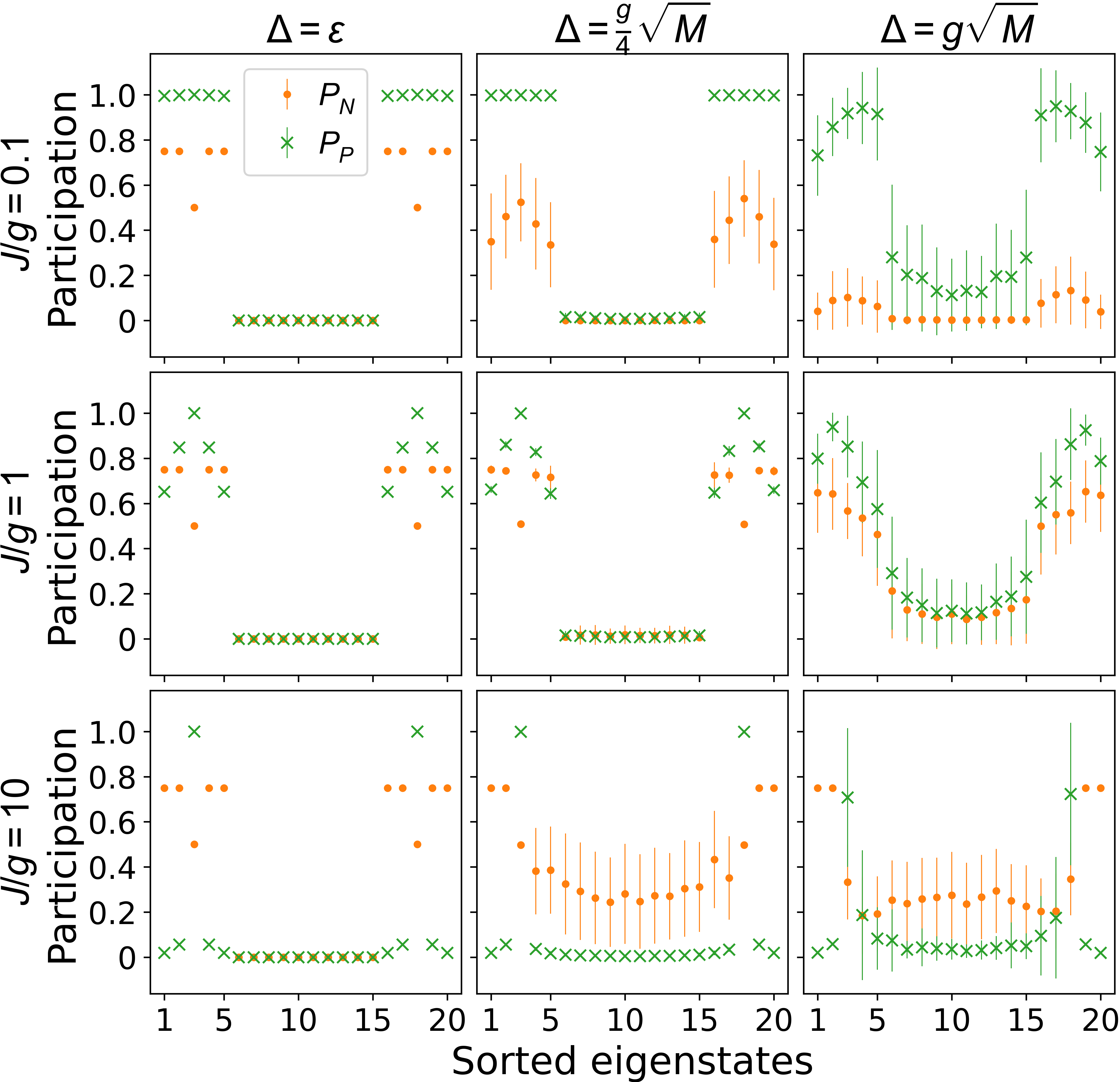}
    \caption{Effects of an increasing disorder, $\Delta$, on the localization and polaritonicity of the eigenstates for 3 different $J/g$ values. For large $J/g$ there remains two polaritonic states located in the middle of the polariton bands. Mean of 100 random realizations; error bars are one standard deviation. $N=5$, $M=3$; $g$, $\kappa$, $\gamma$ same as those in Figure \ref{fig:cavpumpall}}.
    \label{fig:5x3P}
\end{figure}

We utilize the Effective Hamiltonian approach to study medium and large TCHM systems and are especially concerned with the influence of the spectral disorder on the polaritonic and localization properties of the model wavefunctions. The Figure \ref{fig:5x3P} shows the energy spectra of an $N=5, M=3$ system for various regimes of cavity QED to photon hopping ratios. For the vanishing disorder, we clearly see the three components of the spectrum illustrated in Figure \ref{fig:nondisordered-eigenspectrum}: two polaritonic bands with $N$ states and a subradiant band with $N(M-1)$ degenerate states. When cavity QED interaction is significant, a band gap opens between the polariton bands and the subradiant states. With an increasing spectral disorder, the band gap closes and the subradiant state degeneracy is lifted.

\begin{figure}[htb]
    \centering
    \includegraphics[width=\linewidth,angle=0]{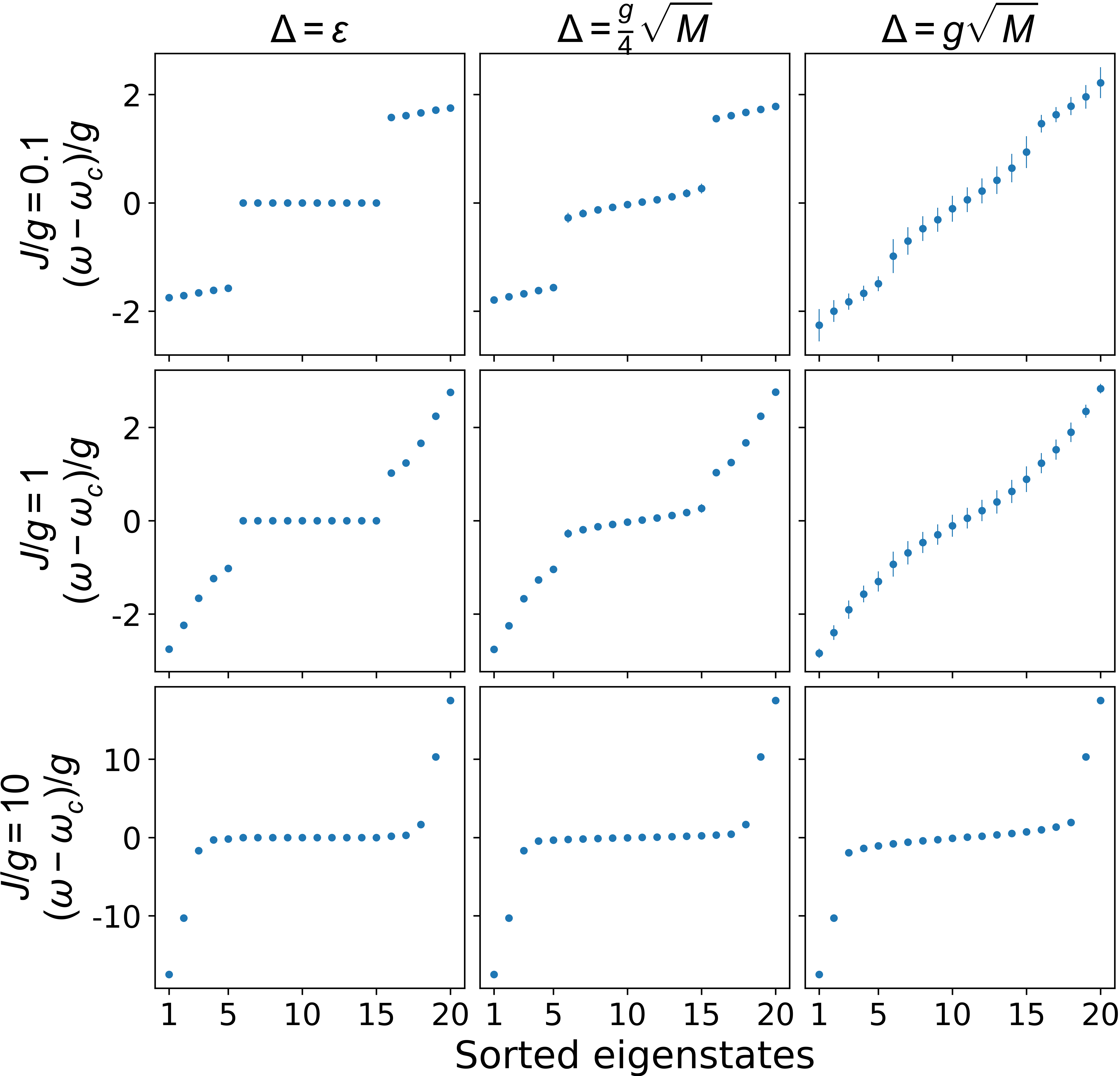}
    \caption{Effects of an increasing disorder, $\Delta$, on the energy eigenspectrum for 3 different $J/g$ values. Mean of 100 random realizations; error bars are one standard deviation. $N=5$, $M=3$; $g$, $\kappa$, $\gamma$ same as those in Figure \ref{fig:cavpumpall}}.
    \label{fig:5x3E}
\end{figure}

Further analysis is provided by the corresponding $P_N$ and $P_P$ values shown in Figure \ref{fig:5x3P}. Here, we observe that the polaritonic properties of the highly hybridized states in the polariton bands and the emitter-like states in the subradiant band shift significantly only for the high levels of spectral disorder, defined by the typical cavity protection cutoff $\Delta=g\sqrt{M}$. The increasing localization trend (decreasing $P_N$) for most polariton band states with an increasing disorder is evident for all sets of parameters. In contrast, the subradiant states gain a cavity component with an increasing disorder and become more hybridized and delocalized.

While the polaritonicity of the lower (as well as the upper) polariton band reduces for most states with an increasing $J/g$ ratio, the middle state of the band remains highly polaritonic. This state, labeled the most polaritonic state (MPS) in our study, shows that even the systems with high hopping ratio can serve as testbeds for polaritonic physics explorations.

\section{A proposed implementation}
\begin{figure}
    \centering
    \includegraphics[width = 0.8\linewidth]{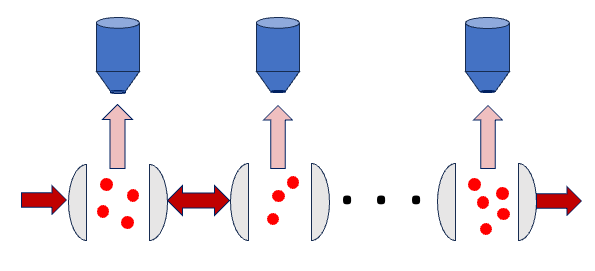}
    \caption{An outline of a potential implementation of a polaritonic simulator for condensed matter systems. The excitation photon (left, red arrow) enters a mirror cavity (grey semi-ovals). The cavity is connected to a number of emitters (red circles). The cavities are coupled to each other with another two-way coupling constant (red double-headed arrows). After excitation, each cavity has a probability of emitting a photon (red vertical arrow) at a rate $\kappa$ which then gets collected by an objective (blue) sent to a detector.}
    \label{fig:prop}
\end{figure}

A physical implementation of the described coupled cavity array system using atoms trapped in optical cavities is depicted in Figure \ref{fig:prop}.
In such a system, the excitation photon would come from a narrow-band tunable laser with low enough power to only target the first excited rung of the Jaynes-Cumming ladder which is what is studied in this work. The ability to collect photons from each cavity will give information about state transfer and localization properties.

\newpage